\documentclass[sigconf, nonacm]{acmart}

\usepackage{listings}
\usepackage[ruled,vlined,linesnumbered]{algorithm2e}
\usepackage[most]{tcolorbox}
\usepackage{makecell}
\usepackage{pgfplots}

\pgfplotsset{compat=1.18}
\SetKwBlock{StageInput}{Stage 1: Generate Test Inputs}{}
\SetKwBlock{StageBuild}{Stage 2: Construct SQL Variants}{}
\SetKwBlock{StageCheck}{Stage 3: Compare Results}{}
\newcounter{finding}
\newcommand{\finding}[1]{
  \refstepcounter{finding}
  \begin{tcolorbox}[
    colback=gray!3,
    colframe=black!60,
    boxrule=0.3pt,
    arc=0.5mm,
    left=3pt,
    right=3pt,
    top=3pt,
    bottom=3pt,
    before skip=5pt,
    after skip=5pt
  ]
  \textbf{Finding~\thefinding.} #1
  \end{tcolorbox}
}

\begin{document}
\title{Detecting DBMS Bugs by Constructing Equivalent Representations of Intermediate Query Results}

\author{Xiaoxu Niu}
\authornote{These authors have contributed equally to this work.}
\affiliation{
  \institution{Wuhan University}
  \department{School of Computer Science}
  \city{Wuhan}
  \country{China}
}
\email{xiaoxuniu@whu.edu.cn}

\author{Gong Chen}
\authornotemark[1]
\affiliation{
  \institution{Wuhan University}
  \department{School of Computer Science}
  \city{Wuhan}
  \country{China}
}
\email{chengongcg@whu.edu.cn}

\author{Jinfu Chen}
\affiliation{
  \institution{Wuhan University}
  \department{School of Computer Science}
  \city{Wuhan}
  \country{China}
}
\email{jinfuchen@whu.edu.cn}

\author{Xiaoyuan Xie}
\authornote{Corresponding author.}
\affiliation{
  \institution{Wuhan University}
  \department{School of Computer Science}
  \city{Wuhan}
  \country{China}
}
\email{xxie@whu.edu.cn}

\begin{abstract}
Database Management Systems (DBMSs) support multiple SQL mechanisms for representing intermediate query results, including VIEWs, Common Table Expressions (CTEs), and Temporary Tables (TEMPTs). When these mechanisms are used to represent the same intermediate query result, the corresponding queries are expected to produce consistent results. However, we observe that such queries can return inconsistent results, indicating potential DBMS logic bugs. Existing approaches for detecting DBMS logic bugs have never explored result consistency across such equivalent representations.

In this paper, we propose ERIQ, a novel testing approach for detecting DBMS logic bugs from the perspective of checking result consistency across \underline{E}quivalent \underline{R}epresentations of \underline{I}ntermediate \underline{Q}uery Results. ERIQ constructs SQL variants using a VIEW, a CTE, or a TEMPT to represent the same intermediate query result, executes these variants, and compares their returned results. We evaluated ERIQ on four widely used open-source DBMSs: MySQL, MariaDB, Percona, and OceanBase. In total, ERIQ detected 64 bugs, 63 of which were confirmed by developers, and two have been fixed. Among the confirmed bugs, 54 were unique and previously unknown logic bugs, and one was a documentation issue.
\end{abstract}

\maketitle

\section{Introduction}\label{sec1}

Database Management Systems (DBMSs) provide various SQL mechanisms for representing intermediate query results, including VIEWs, Common Table Expressions (CTEs), and Temporary Tables (TEMPTs) \cite{sql_standard, mysql_view, mysql_cte, mysql_temp_table}. 
These mechanisms are widely used in database applications to simplify the construction of complex queries and improve query readability: VIEWs define reusable named queries, CTEs provide temporary result sets that can be referenced within a single statement, and TEMPTs store query results in temporary tables that can be reused by subsequent statements \cite{mysql_view,mysql_cte,mysql_temp_table}. 
When different mechanisms are used to represent the same intermediate query result, the corresponding queries are expected to produce consistent results. 
\textbf{However, we observe that such queries can return inconsistent results, as illustrated in Section~\ref{sec2.2}.} Such inconsistencies can indicate potential DBMS bugs, which may be exposed when the corresponding queries are processed through different paths within the DBMS.

Many approaches have been proposed for detecting DBMS logic bugs. \textbf{However, to the best of our knowledge, they have never explored the perspective of detecting such bugs by checking result consistency across equivalent representations of intermediate query results.} For example, NoREC~\cite{norec} compares the results of optimized and unoptimized queries; TLP~\cite{tlp} partitions queries based on ternary logic; PQS~\cite{pqs} checks whether a selected pivot row appears in the result of a generated query; Pinolo~\cite{pinolo} checks result containment between synthesized queries and a seed query; EET~\cite{eet} and CODDTest~\cite{coddtest} apply equivalent transformations at the expression level; Radar~\cite{radar} compares query results across databases with the same data but different metadata; and EDC~\cite{edc} checks query consistency using precomputed results of data operation expressions.

In this paper, we propose ERIQ, a novel testing approach for detecting DBMS logic bugs from the perspective of checking result consistency across \underline{E}quivalent \underline{R}epresentations of \underline{I}ntermediate \underline{Q}uery Results. ERIQ constructs corresponding SQL variants that use a VIEW, a CTE, or a TEMPT to represent the same intermediate query result, executes these variants, and compares their returned results. If the results are inconsistent, ERIQ reports the inconsistency as a bug.

We implemented ERIQ and evaluated it on four widely used open-source DBMSs: MySQL, MariaDB, Percona, and OceanBase. In total, ERIQ detected 64 bugs, including 29 in MySQL, 18 in MariaDB, 13 in Percona, and 4 in OceanBase. Developers confirmed 63 of them, including 54 unique and previously unknown logic bugs, 8 duplicates of other bugs reported by ERIQ, and one documentation issue. Among the 54 unique logic bugs, two have been fixed. We further compared ERIQ with four state-of-the-art DBMS logic bug detection approaches: EDC~\cite{edc}, Radar~\cite{radar}, EET~\cite{eet}, and TLP~\cite{tlp}. Among these 54 unique logic bugs, the best-performing existing approach could detect only 9 of them, showing that ERIQ detects many bugs missed by each existing approach and thus provides complementary bug detection capability. Under the 24-hour testing budget on each supported DBMS, ERIQ detected 54 logic bugs, compared with 16 detected by the best-performing existing approach, demonstrating its stronger bug detection capability.

This paper makes the following contributions:
\begin{itemize}
\item We introduce a new perspective for detecting DBMS logic bugs by checking result consistency across equivalent representations of intermediate query results, which has never been explored by existing methods.
\item Based on this new perspective, we propose and implement ERIQ, a novel testing approach that constructs SQL variants using a VIEW, a CTE, or a TEMPT to represent the same intermediate query result and detects DBMS logic bugs by checking result consistency across these variants.
\item We conduct a comprehensive experimental study of ERIQ on four widely used open-source DBMSs, detecting 54 unique and previously unknown logic bugs and demonstrating that ERIQ complements existing approaches.
\end{itemize}

The rest of this paper is organized as follows. Section~\ref{sec2} introduces the background and motivation. Section~\ref{sec3} describes the proposed approach. Section~\ref{sec4} presents the experimental evaluation. Section~\ref{sec5} discusses false positive mitigation, a documentation issue, and limitations. Section~\ref{sec6} reviews related work. Finally, Section~\ref{sec7} concludes the paper.

\section{Background and Motivation}\label{sec2}

This section introduces different representations of intermediate query results and their representative processing paths, presents a real-world logic bug that motivates our work, and derives the key insight underlying our approach.

\subsection{Representations of Intermediate Query Results}\label{sec2.1}

\begin{figure}[t] %[htbp]
\centering
\includegraphics[width=\columnwidth]{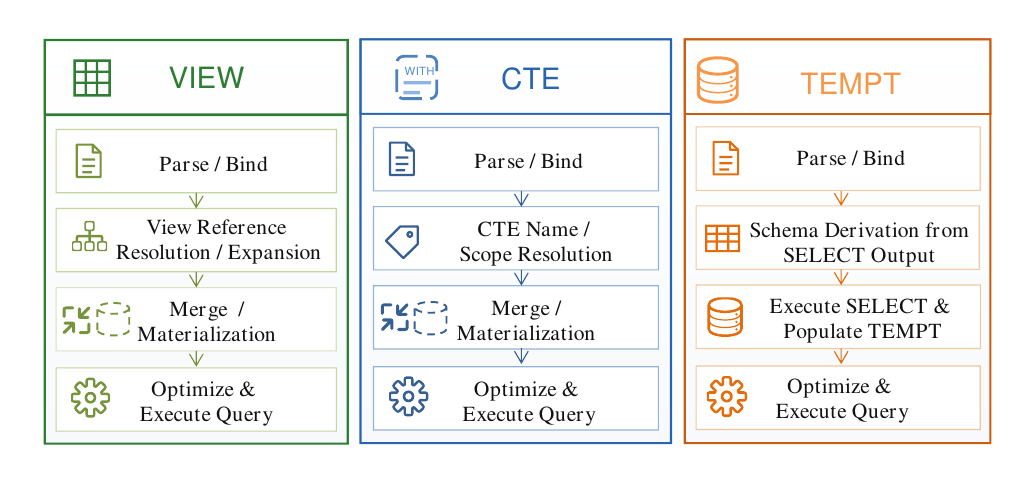}
\caption{Simplified processing paths for VIEW, CTE, and TEMPT in MySQL.}
\label{fig1}
\vspace{-4mm}
\end{figure}

VIEWs, CTEs, and TEMPTs provide different ways to represent intermediate query results in SQL. Using MySQL as an example, \autoref{fig1} illustrates their simplified representative processing paths. 
\textbf{VIEWs} define reusable named queries whose results can be referenced by other queries~\cite{mysql_view}. Processing a VIEW reference may involve resolving and expanding its definition, after which the optimizer may merge the VIEW's defining query into the referencing query or use materialization during query execution~\cite{mysql_derived_opt}. \textbf{CTEs} provide temporary result sets that can be referenced within a single statement~\cite{mysql_cte}. Processing a CTE may similarly involve resolving its statement-scoped name and selecting between merging and materialization during query optimization~\cite{mysql_derived_opt}. \textbf{TEMPTs} store query results in temporary tables that can be reused by subsequent statements~\cite{mysql_temp_table}. For a TEMPT created using \texttt{CREATE TEMPORARY TABLE ... AS SELECT}, the DBMS derives the temporary-table schema from the output of the defining query, executes that query to populate the temporary table, and subsequently optimizes and executes a separate query over the temporary table~\cite{mysql_ctas}.
Although these mechanisms differ in syntax, scope, lifetime, and materialization behavior, they can be used to represent the same intermediate query result. In this paper, equivalence refers to the semantic equivalence of the represented intermediate query result, rather than identical internal representations, optimization strategies, or execution paths. When they are used in this way, the corresponding queries are expected to return consistent results.

\begin{figure*}[htbp]
  \centering
  \includegraphics[width=0.93\textwidth]{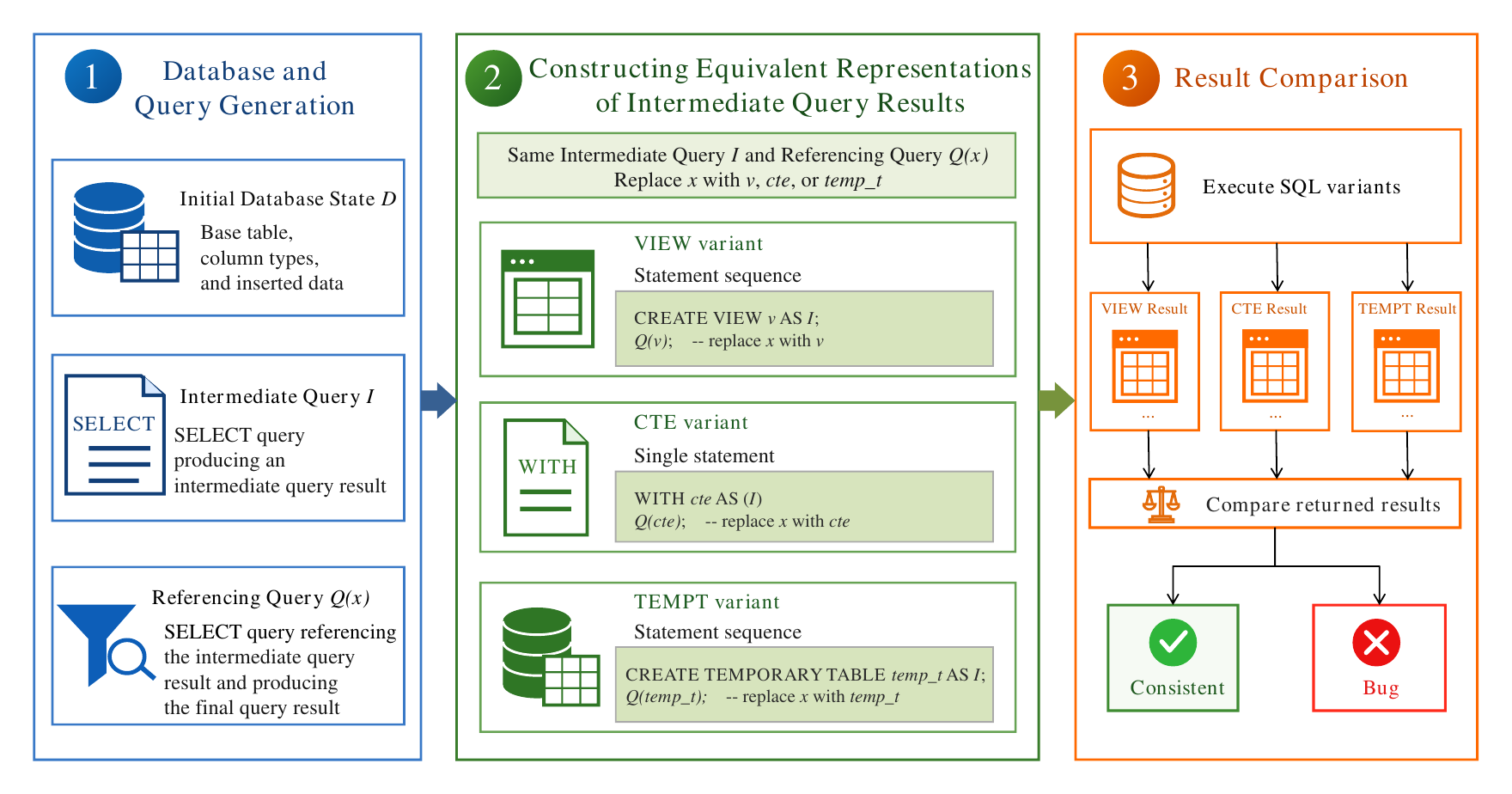}
  \caption{Overview of ERIQ.}
  \label{fig2}
  \vspace{-2mm}
\end{figure*}

\subsection{Motivating Example}\label{sec2.2}

\textbf{Listing~\ref{lst1} presents \href{https://jira.mariadb.org/browse/MDEV-40143}{MDEV\#40143}, a real logic bug in MariaDB detected in our study and confirmed by the developers.}
The test case first creates a table \texttt{t0} with an integer column and a spatial column, and inserts three rows. It then constructs three SQL variants that use a VIEW, a CTE, and a TEMPT, respectively, to represent the same intermediate query result. In all three variants, the intermediate query result is produced by the same \texttt{SELECT} query, which joins two instances of \texttt{t0} using a \texttt{RIGHT JOIN} with a null-safe equality predicate (\texttt{<=>}), filters rows using \texttt{ST\_Dimension(b.c1) >= 0}, and computes \texttt{ST\_IsValid(a.c1)} as \texttt{vc\_0}. Each variant then selects \texttt{vc\_0} from the corresponding representation and applies the same predicate \texttt{vc\_0 > 0}.

\begin{lstlisting}[
 float=htbp, 
caption={MDEV\#40143. A real logic bug in MariaDB.},label={lst1},language=SQL,frame=single, basicstyle=\footnotesize\ttfamily, escapeinside={(@}{@)},
aboveskip=-1em,
  belowskip=-1em
]
CREATE TABLE t0 (c0 INT, c1 POINT);
INSERT INTO t0 VALUES (-4, Point(0,0));
INSERT INTO t0 VALUES (6, NULL);
INSERT INTO t0 VALUES (9, Point(1,1));

-- VIEW variant
CREATE VIEW v0 AS SELECT ST_IsValid(a.c1) AS vc_0
  FROM t0 AS a RIGHT JOIN t0 AS b ON a.c0 <=> b.c0
  WHERE ST_Dimension(b.c1) >= 0;
SELECT vc_0 FROM v0 WHERE vc_0 > 0;
(@\textcolor{red}{-- Returns: 1}@)

-- CTE variant
WITH CTE AS (SELECT ST_IsValid(a.c1) AS vc_0
  FROM t0 AS a RIGHT JOIN t0 AS b ON a.c0 <=> b.c0
  WHERE ST_Dimension(b.c1) >= 0)
SELECT vc_0 FROM CTE WHERE vc_0 > 0;
(@\textcolor{red}{-- Returns: NULL}@)

-- TEMPT variant
CREATE TEMPORARY TABLE temp_t AS 
  SELECT ST_IsValid(a.c1) AS vc_0 
  FROM t0 AS a RIGHT JOIN t0 AS b ON a.c0 <=> b.c0
  WHERE ST_Dimension(b.c1) >= 0;
SELECT vc_0 FROM temp_t WHERE vc_0 > 0;
(@\textcolor{red}{-- Returns: 1, 1}@)
\end{lstlisting}

According to the SQL semantics, the three variants should return the same result.
\textbf{However, MariaDB produces inconsistent results across the three variants}: the VIEW variant returns only one row with value~\texttt{1}, the CTE variant returns one row with value~\texttt{NULL}, and the TEMPT variant returns two rows with value~\texttt{1}. MariaDB developers confirmed that the expected result is two rows with value~\texttt{1}: \texttt{ST\_Dimension(b.c1) >= 0} retains two joined rows, and \texttt{ST\_IsValid(a.c1)} evaluates to~\texttt{1} for both, so both rows satisfy \texttt{vc\_0 > 0}. Therefore, the TEMPT variant returns the expected result, whereas the VIEW and CTE variants produce incorrect results.

This example demonstrates that DBMS logic bugs can manifest as inconsistent results across queries constructed using equivalent representations of the same intermediate query result. MariaDB developers have fixed this bug
(\url{https://jira.mariadb.org/browse/MDEV-40143}), and its root cause is further analyzed in Section~\ref{sec4.5}. Importantly, such logic bugs may silently produce incorrect query results without causing crashes or explicit errors, potentially propagating incorrect data to downstream applications.

\subsection{Key Insight}\label{sec2.3}
\textbf{This observation motivates a new perspective for detecting DBMS logic bugs by checking result consistency across equivalent representations of the same intermediate query result.} However, existing DBMS logic bug detection approaches have never explored this perspective. For example, existing approaches expose bugs by transforming, partitioning, or constructing queries~\cite{norec,tlp,pqs,pinolo}, transforming query expressions~\cite{eet,coddtest}, varying database metadata~\cite{radar}, or using precomputed results of data operation expressions~\cite{edc}.

The rationale behind this perspective follows from the observation in Section~\ref{sec2.1}: queries constructed using equivalent representations of the same intermediate query result may be processed through different paths within the DBMS. Consequently, an underlying DBMS logic bug can affect the corresponding queries differently, resulting in inconsistent query results, as demonstrated by the real-world bug presented in Section~\ref{sec2.2}. Therefore, constructing such equivalent representations and comparing the corresponding query results provides an effective approach to detecting DBMS logic bugs.

\section{Approach}\label{sec3}
In this paper, we propose ERIQ, a novel testing approach for detecting DBMS logic bugs by checking result consistency across \underline{E}quivalent \underline{R}epresentations of \underline{I}ntermediate \underline{Q}uery Results. We first provide an overview of ERIQ and then describe database and query generation, the construction of equivalent representations of intermediate query results, and result comparison.

\subsection{Overview of ERIQ}\label{sec3.1}
\autoref{fig2} provides an overview of ERIQ, which consists of three stages. 
\ding{172} \textbf{Database and Query Generation.}
ERIQ first generates an initial database state $D$, including a base table, its column types, and inserted data. It then generates an intermediate query $I$, which is a \texttt{SELECT} query that produces an intermediate query result from the generated database. Based on the names and types of the columns produced by $I$, ERIQ further generates a referencing query $Q(x)$, which references the intermediate query result through $x$ and produces the final query result, where $x$ is a placeholder for the name of the VIEW, CTE, or TEMPT used to represent the intermediate query result.
\ding{173} \textbf{Constructing Equivalent Representations of Intermediate Query Results.}
ERIQ constructs SQL variants that use a VIEW, a CTE, or a TEMPT to represent the result produced by the same intermediate query $I$. For each representation, ERIQ replaces $x$ in the same referencing query $Q(x)$ with the name of the corresponding VIEW, CTE, or TEMPT. In this way, the SQL variants use equivalent representations of the same intermediate query result, while the resulting referencing queries differ only in the relation name used to reference each representation.
\ding{174} \textbf{Result Comparison.}
ERIQ executes the constructed SQL variants and compares their returned results. If the results are inconsistent, ERIQ reports the inconsistency as a bug.

Algorithm~\ref{alg:eriq} summarizes the workflow of ERIQ. Within a given time budget, ERIQ repeatedly generates an initial database state $D$, an intermediate query $I$, and a referencing query $Q(x)$. It then constructs the corresponding SQL variants, executes them, and compares their returned results. An inconsistency report is generated when the compared variants return different results.

\begin{algorithm}[htbp]
\caption{The workflow of ERIQ}
\label{alg:eriq}
\DontPrintSemicolon
\KwIn{target DBMS $M$, time budget $T$}
\KwOut{inconsistency reports $Reports$}

$Reports \leftarrow \emptyset$\;

\While{the time budget $T$ is not exhausted}{

\StageInput{
    $D \leftarrow \textit{generateDatabase}(M)$\;
    $I \leftarrow \textit{generateIntermediateQuery}(D,M)$\;
    $Q(x) \leftarrow \textit{generateReferencingQuery}(I,M)$\;
}

\StageBuild{
    $Variants \leftarrow \emptyset$\;
    \ForEach{$m \in \{\textit{VIEW},\textit{CTE},\textit{TEMPT}\}$}{
        $s \leftarrow \textit{buildVariant}(I,Q,m,M)$\;
        \If{$s$ can be legally constructed}{
            add $(m,s)$ to $Variants$\;
        }
    }
}

\StageCheck{
    $Results \leftarrow \emptyset$\;

    \ForEach{$(m,s) \in Variants$}{
        $r \leftarrow \textit{execute}(s,D,M)$\;
        \If{execution succeeds and returns $r$}{
            $nr \leftarrow \textit{normalize}(r)$\;
            add $(m,nr)$ to $Results$\;
        }
    }

    \If{$|\mathit{Results}| \geq 2$ and all constructed variants return results \textbf{and}
        $Results$ are inconsistent}{
        add an inconsistency report to $Reports$\;
    }
}

$\textit{cleanup}(D,M)$\;
}

\Return{$Reports$}\;
\end{algorithm}

\subsection{Database and Query Generation}\label{sec3.2}
This subsection describes how ERIQ generates the three core components of each test case: the initial database state $D$, the intermediate query $I$, and the referencing query $Q(x)$. The initial database state $D$ provides the base-table schema and data used for testing. The intermediate query $I$ is a \texttt{SELECT} query that produces an intermediate query result from $D$, while the referencing query $Q(x)$ references this intermediate query result and produces the final query result.

\textbf{Database Generation.}
ERIQ first generates an initial database state $D$ consisting of a base-table schema and inserted data. To exercise type-related query-processing logic, ERIQ selects column types from a broad range of data types supported by the target DBMS, including integer, floating-point, fixed-point, bit-value, temporal, string, JSON, and spatial types. After constructing the table schema, ERIQ generates type-compatible values and inserts them into the table. The generated data include ordinary random values, special and boundary-oriented values such as \texttt{NULL}, empty strings, zeros, negative numbers, large numeric values, and boundary temporal values, as well as representative JSON and spatial values.

To improve the validity and diversity of generated SQL statements, ERIQ adopts a type-aware generation strategy for expressions and predicates. Rather than randomly combining columns, constants, operators, and functions, ERIQ considers their type compatibility when constructing expressions. For example, numeric columns are used in arithmetic and comparison expressions, string columns in pattern-matching and length expressions, and temporal columns in date/time extraction and interval-arithmetic expressions. JSON, spatial, and bit-value columns are similarly combined with compatible built-in functions and operators. This strategy reduces type-incompatible combinations and facilitates the generation of valid and structurally diverse SQL statements.

\textbf{Intermediate Query Generation.}
ERIQ generates an intermediate query $I$, which is a \texttt{SELECT} query that produces an intermediate query result from the generated database. To exercise diverse DBMS query-processing logic, ERIQ generates intermediate queries with diverse structures, including projection, \texttt{WHERE} filtering, \texttt{ORDER BY} and \texttt{LIMIT}, \texttt{DISTINCT}, \texttt{GROUP BY} and aggregation, \texttt{UNION} and \texttt{UNION ALL}, inner and outer joins, \texttt{IN}/\texttt{NOT IN} subqueries, \texttt{EXISTS}/\texttt{NOT EXISTS} subqueries, scalar subqueries, derived tables, joins over derived tables, window functions, and aggregate queries over derived results. These structures exercise DBMS logic related to name resolution, type inference, predicate evaluation, join processing, subquery rewriting, set operations, aggregation, and window-function evaluation.

ERIQ assigns stable aliases, such as \texttt{vc\_0} and \texttt{vc\_1}, to the output expressions of $I$. These aliases provide consistent column names for the intermediate query result when it is represented using a VIEW, a CTE, or a TEMPT.

\textbf{Referencing Query Generation.}
After generating $I$, ERIQ generates a referencing query $Q(x)$. In $Q(x)$, $x$ is a placeholder for the name of the VIEW, CTE, or TEMPT used to represent the intermediate query result. $Q(x)$ references the intermediate query result through $x$ and produces the final query result. Based on the names and types of the columns produced by $I$, ERIQ generates $Q(x)$ with diverse structures, including filtering, aggregation, joins, subqueries, window functions, and complex expressions and predicates. $Q(x)$ may also reference $x$ multiple times within the same statement.

To obtain information about the columns produced by $I$, ERIQ first creates the VIEW representation of $I$ and obtains its column names and types. This information is used only to generate a type-valid referencing query; the result returned by this VIEW is not used as an expected result during result comparison.

When generating the intermediate query and the referencing query, ERIQ applies two additional constraints to reduce inconsistencies caused by nondeterminism. First, nondeterministic and time-dependent functions, such as \texttt{RAND()} and \texttt{NOW()}, are excluded from both the intermediate query and the referencing query because different executions may legitimately return different values. Second, when \texttt{ORDER BY} is combined with \texttt{LIMIT}, ERIQ adds additional ordering expressions to reduce ties among rows with equal ordering keys, thereby reducing inconsistencies caused by nondeterministic row selection.

\subsection{Constructing Equivalent Representations of Intermediate Query Results} \label{sec3.3}

Given the intermediate query $I$ and referencing query $Q(x)$, ERIQ constructs SQL variants that use a VIEW, a CTE, or a TEMPT to represent the same intermediate query result. For clarity, let $S_V(I,Q)$, $S_C(I,Q)$, and $S_T(I,Q)$ denote the VIEW, CTE, and TEMPT variants, respectively. Conceptually, they are constructed as follows:

\begin{equation}
\begin{aligned}
S_V(I,Q) ={}&
[\,\texttt{CREATE VIEW v AS } I;\ Q(\texttt{v})\,],\\
S_C(I,Q) ={}&
[\,\texttt{WITH cte AS (}I\texttt{) }\ Q(\texttt{cte})\,],\\
S_T(I,Q) ={}&
[\,\texttt{CREATE TEMPORARY TABLE}\\
&\quad \texttt{temp\_t AS } I;\ Q(\texttt{temp\_t})\,].
\end{aligned}
\label{eq:variants}
\end{equation}

Here, $Q(\texttt{v})$, $Q(\texttt{cte})$, and $Q(\texttt{temp\_t})$ denote the same generated referencing query $Q(x)$, with $x$ replaced by \texttt{v}, \texttt{cte}, and \texttt{temp\_t}, respectively. The brackets denote the SQL statement or statement sequence constituting one SQL variant. In the VIEW variant, the result produced by $I$ is represented using VIEW \texttt{v}, which is referenced by $Q(\texttt{v})$. In the CTE variant, the result produced by $I$ is represented using CTE \texttt{cte}, which is referenced by $Q(\texttt{cte})$ within the same SQL statement. In the TEMPT variant, the result produced by $I$ is stored in \texttt{temp\_t}, which is referenced by $Q(\texttt{temp\_t})$.

\textbf{The three variants are constructed from the same intermediate query $I$ and the same generated referencing query $Q(x)$.} Their intended difference is the mechanism used to represent the intermediate query result. ERIQ therefore does not compare the result produced by $I$ in isolation. Instead, it compares the results returned by the SQL variants after the corresponding representation is referenced by $Q(x)$.

When all three variants can be legally constructed and successfully return query results, their expected result consistency can be expressed as:

\begin{equation}
R(S_V(I,Q),D)
\equiv
R(S_C(I,Q),D)
\equiv
R(S_T(I,Q),D),
\label{eq:result-consistency}
\end{equation}

where $R(S,D)$ denotes the query result returned by SQL variant $S$ under database state $D$, and $\equiv$ denotes result equivalence according to the comparison procedure described in Section~\ref{sec3.4}.

Not every representation can be legally constructed for every referencing query. In particular, when $Q(x)$ references the intermediate query result multiple times within the same statement, a TEMPT variant may not be legally constructible on some target DBMSs because these systems prohibit multiple references to the same temporary table within a single statement~\cite{mysql_temp_limits}. In such cases, ERIQ excludes the TEMPT variant and compares the VIEW and CTE variants if both successfully return query results.

\subsection{Result Comparison}\label{sec3.4}
After constructing the SQL variants, ERIQ executes all legally constructed variants. Result comparison is performed only when at least two variants are constructed and all constructed variants successfully return query results. If any constructed variant fails to execute or does not return a query result, ERIQ discards the test case rather than treating the execution failure as a bug. Before comparison, ERIQ normalizes returned values to reduce spurious inconsistencies caused by representational and minor numerical differences. For representational differences, ERIQ converts values returned in different forms into stable comparable representations. For example, byte-string and byte-array values are converted into textual, hexadecimal, or numeric representations depending on their contents, while decimal values are canonicalized so that numerically equivalent forms have the same representation. For minor numerical differences, ERIQ normalizes finite floating-point values using a fixed-precision scientific-notation representation. This normalization addresses minor variations in floating-point results introduced by numerical precision and data-type conversions. Such variations can arise when approximate floating-point values are materialized into table columns, where data-type conversion may occur~\cite{mysql_ctas,mysql_float}.

ERIQ compares the normalized results under multiset semantics. Two results are considered equivalent if and only if they contain the same normalized rows with identical multiplicities. This comparison captures both differences in returned values and differences in the number of duplicate rows. For example, a result containing one row with value~\texttt{1} is different from a result containing two rows with value~\texttt{1}, even though both contain the same distinct value. Differences in row order alone are not treated as inconsistencies. However, row ordering may affect which rows are returned when \texttt{ORDER BY} is combined with \texttt{LIMIT}. As described in Section~\ref{sec3.2}, ERIQ adds additional ordering expressions to reduce ties among rows with equal ordering keys, thereby reducing inconsistencies caused by nondeterministic row selection.

ERIQ classifies inconsistencies according to the number of compared variants and the relationships among their returned results. When all three variants are compared, an inconsistency is classified into one of four patterns: $\mathtt{VIEW=TEMPT\ne CTE}$, $\mathtt{VIEW=CTE\ne TEMPT}$, $\mathtt{CTE=TEMPT\ne VIEW}$, and \texttt{all\_diff}, where the first three indicate that one variant differs from the equal results of the other two, and \texttt{all\_diff} indicates that all three results are pairwise different. When the TEMPT variant cannot be legally constructed, ERIQ compares the VIEW and CTE variants; if their results differ, the inconsistency is recorded as $\mathtt{VIEW\ne CTE}$\texttt{(TEMPT\_SKIP)}. ERIQ therefore records five inconsistency patterns.

For each detected inconsistency, ERIQ generates an inconsistency report containing the database schema, inserted data, intermediate query $I$, referencing query $Q(x)$, constructed SQL variants, returned results and result-row counts, and the inconsistency pattern. ERIQ reports the inconsistency as a bug. 

\section{Evaluation}\label{sec4}
This section evaluates the effectiveness of ERIQ in detecting DBMS bugs. Specifically, we investigate the following research questions:

\begin{itemize}
\item \textbf{RQ1: How effective is ERIQ in detecting DBMS bugs?} In this RQ, we analyze the detected bugs and their developer-confirmed status to evaluate ERIQ's overall bug detection effectiveness.

\item \textbf{RQ2: How does ERIQ perform compared with existing approaches?} In this RQ, we compare ERIQ with existing approaches to evaluate its complementary bug detection capability and its bug detection capability under the same testing budget.

\item \textbf{RQ3: How do the three representations of intermediate query results contribute to ERIQ's bug detection?} In this RQ, we compare different combinations of VIEW, CTE, and TEMPT representations to analyze whether these representations provide complementary bug detection capability.

\item \textbf{RQ4: What characteristics do the bugs detected by ERIQ exhibit?} In this RQ, we analyze the SQL features involved in the detected bugs and examine representative root causes to understand the types of DBMS defects exposed by ERIQ.
\end{itemize}

\subsection{Experimental Setup}\label{sec4.1}
\textbf{Target DBMSs.} We selected 4 widely used open-source DBMSs: MySQL~\cite{mysql}, MariaDB~\cite{mariadb}, Percona~\cite{percona}, and OceanBase~\cite{oceanbase}. We tested the latest release versions available when the experiments began: MySQL 9.6.0, MariaDB 12.2.2, Percona 8.0.45-36, and OceanBase 4.4.2.0. These DBMSs have been extensively tested by many existing approaches~\cite{edc,dynsql,dqe,dqp,puppy,wingfuzz}. Therefore, finding new bugs in them is challenging.

\textbf{Baselines.}
We selected 4 state-of-the-art DBMS logic bug detection approaches as baselines: EDC~\cite{edc}, Radar~\cite{radar}, EET~\cite{eet}, and TLP~\cite{tlp}. These approaches have demonstrated their effectiveness by uncovering real-world DBMS logic bugs and represent different bug detection perspectives. EDC was selected as the primary baseline because it supports all target DBMSs in our evaluation. Radar, EET, and TLP were additionally included because they were used as baselines in EDC's evaluation. These baselines enable us to evaluate whether ERIQ provides complementary bug detection capability by introducing a new testing perspective.

\textbf{Experimental infrastructure.} We conducted all experiments on a server equipped with an Intel Xeon Gold 6254 processor (3.10 GHz, 18 cores, and 36 hardware threads), 64 GB of RAM, and approximately 1 TB of local storage. The server ran 64-bit Ubuntu 20.04.3 LTS. Each DBMS was deployed in a separate Docker container created from its official image and run with its default configuration.

\subsection{RQ1: Effectiveness of ERIQ}\label{sec4.2}

To answer RQ1, we evaluated ERIQ's effectiveness based on the bugs it detected and their developer-confirmed status. We applied ERIQ to the 4 target DBMSs and ran it on each DBMS for 24 hours. Across all experiments, ERIQ generated 2,316 inconsistency reports. Because multiple reports may correspond to the same underlying bug, we minimized and internally deduplicated the reports before submitting them to the developers.

Specifically, we first grouped the reports according to the structures of their intermediate queries and referencing queries to identify structurally similar cases. We then selected representative test cases from each group and minimized them using C-Reduce~\cite{creduce} together with manual simplification, while ensuring that the result inconsistency remained reproducible. Finally, we further deduplicated the minimized test cases by comparing their suspected root causes, involved functions, SQL keywords, and query structures. This process took approximately two weeks, after which we obtained 64 bugs: 29 in MySQL, 18 in MariaDB, 13 in Percona, and 4 in OceanBase.

\textbf{Bug Status.}
We reported the 64 bugs to their respective DBMS communities and tracked their status. \autoref{tab1} summarizes the developer responses at the time of writing. Developers confirmed 63 of the 64 bugs: 54 were unique and previously unknown logic bugs, 8 were classified as duplicates of other bugs reported by ERIQ during our evaluation, and 1 was classified as a documentation issue. We discuss the documentation issue in Section~\ref{sec5}. The remaining report was still under investigation. Among the 54 unique logic bugs, 2 have been fixed. These results show that ERIQ is effective at detecting real DBMS bugs.

\begin{table}[htbp]
\centering
\caption{Status of bugs detected by ERIQ.}
\label{tab1}
\resizebox{\columnwidth}{!}{
\begin{tabular}{lccccc}
\toprule
\textbf{DBMS} &
\textbf{Reported} &
\multicolumn{2}{c}{\textbf{Logic Bug}} &
\textbf{Documentation Issue} &
\textbf{Investigating} \\
\cmidrule(lr){3-4}
&
&
\textbf{Unique (Fixed)} &
\textbf{Duplicate} &
&
\\
\midrule
MySQL     & 29 & 26 (1) & 2 & 1 & 0 \\
MariaDB   & 18 & 11 (1) & 6 & 0 & 1 \\
Percona   & 13 & 13 (0) & 0 & 0 & 0 \\
OceanBase &  4 &  4 (0) & 0 & 0 & 0 \\
\midrule
\textbf{Total} &
\textbf{64} &
\textbf{54 (2)} &
\textbf{8} &
\textbf{1} &
\textbf{1} \\
\bottomrule
\end{tabular}
}
\end{table}

\textbf{Bug Severity.}
Among the 54 unique logic bugs, 48 were classified as Critical, Serious, Major, or High by developers. Specifically, MySQL developers classified 1 bug as Critical and 24 as Serious; MariaDB developers classified 10 bugs as Major; and Percona developers classified 13 bugs as High. Among the remaining 6 bugs, 1 MySQL bug was classified as Non-critical, 1 MariaDB bug was classified as Minor, and the 4 OceanBase bugs had not been assigned severity classifications. These results indicate that most of the unique logic bugs detected by ERIQ received relatively high severity classifications from the respective DBMS communities.

\finding{ERIQ detected 64 bugs, 63 of which were confirmed by developers, and 2 have been fixed. Among the confirmed bugs, 54 were unique and previously unknown logic bugs, and 1 was a documentation issue. Of the 54 unique logic bugs, 48 received relatively high severity classifications.}

\subsection{RQ2: Comparison with Existing Approaches}\label{sec4.3}
To answer RQ2, we compared ERIQ with the 4 baselines introduced in Section~\ref{sec4.1}: EDC~\cite{edc}, Radar~\cite{radar}, EET~\cite{eet}, and TLP~\cite{tlp}. We conducted two complementary experiments. First, we analyzed the distinctiveness of ERIQ's 54 unique logic bugs by examining whether existing approaches could also detect them. Second, we compared ERIQ and the baselines under the same time budget to evaluate their practical bug detection capability.

\textbf{Distinctiveness of Bugs Detected by ERIQ.}
For each of the 54 unique logic bugs detected by ERIQ, we analyzed whether they could also be detected by existing approaches. Specifically, we started from each minimized test case and constructed corresponding test cases following the relation and construction procedure defined by each approach~\cite{edc,radar,eet,tlp}. A bug was considered detectable by a baseline if the corresponding test case violated the relation checked by that baseline; otherwise, it was considered not detectable by that baseline. Because this analysis evaluates the relations checked by the baselines rather than their released implementations, it is independent of whether those implementations support the corresponding DBMSs. \autoref{tab3} presents the results.

\begin{table}[htbp]
\centering
\caption{Number of ERIQ's 54 unique logic bugs detectable by existing approaches.}
\label{tab3}
\begin{tabular}{lrrrrr}
\toprule
\textbf{DBMS} & \textbf{ERIQ} & \textbf{EDC} & \textbf{Radar} & \textbf{EET} & \textbf{TLP} \\
\midrule
MySQL      & 26  & 4 & 4  & 5  & 6  \\
MariaDB    & 11  & 2 & 1  & 1  & 3  \\
Percona    & 13  & 2  & 1  & 0  & 0  \\
OceanBase  & 4   & 0  & 0  & 0  & 0  \\
\midrule
\textbf{Total} & \textbf{54} & \textbf{8} & \textbf{6} & \textbf{6} & \textbf{9} \\
\bottomrule
\end{tabular}
\end{table}

Among the 54 unique logic bugs detected by ERIQ, EDC, Radar, EET, and TLP could detect 8, 6, 6, and 9 bugs, respectively, corresponding to 14.8\%, 11.1\%, 11.1\%, and 16.7\%. TLP detected the largest number of ERIQ-detected bugs in this analysis, but only 9 of the 54 bugs; the other baselines detected even fewer. For each baseline, at least 45 of the 54 bugs detected by ERIQ were not detected. These results show that ERIQ provides complementary bug detection capability by exposing many bugs that are not detected by existing approaches.

\textbf{Bug Detection Capability under the Same Time Budget.}
We further compared the practical bug detection capability of ERIQ and the baselines under the same time budget. We ran each tool for 24 hours on every target DBMS it supports. For ERIQ, we reused the results reported in Section~\ref{sec4.2}. For each baseline, we used its publicly available implementation and executed it on all supported target DBMSs. All experiments used the same DBMS versions and experimental environment described in Section~\ref{sec4.1}. For each baseline, we minimized and internally deduplicated the generated reports following the general procedure described in Section~\ref{sec4.2}. The detected bugs were reported to the corresponding DBMS communities, and developer responses confirmed that they corresponded to real bugs.

\autoref{tab4} presents the number of logic bugs detected by each tool within 24 hours. Across the 4 target DBMSs, ERIQ detected 54 logic bugs, whereas EDC, Radar, EET, and TLP detected 16, 2, 1, and 6 bugs, respectively. Among the baselines, EDC detected the largest number of bugs and, like ERIQ, was evaluated on all 4 target DBMSs. ERIQ detected 54 logic bugs compared with EDC's 16, more than three times as many under the same time budget. At the time of writing, developers had not classified any of the bugs detected by the baselines as duplicates of the 54 unique logic bugs found by ERIQ. These results demonstrate that ERIQ detects more logic bugs than existing approaches under the same time budget and provide additional evidence that ERIQ complements these baselines.

\begin{table}[htbp]
\centering
\caption{Number of logic bugs detected within 24 hours. ``--'' indicates that the approach does not support the DBMS.}
\label{tab4}
\begin{tabular}{lrrrrr}
\toprule
\textbf{DBMS} & \textbf{ERIQ} & \textbf{EDC} & \textbf{Radar} & \textbf{EET} & \textbf{TLP} \\
\midrule
MySQL      & 26  & 3  & 1  & 1  & 2 \\
MariaDB    & 11  & 3  & 0  & -- & -- \\
Percona    & 13  & 6  & 1  & -- & 4 \\
OceanBase  & 4   & 4  & -- & -- & 0 \\
\midrule
\textbf{Total} & \textbf{54} & \textbf{16} & \textbf{2} & \textbf{1} & \textbf{6} \\
\bottomrule
\end{tabular}
\end{table}

\finding{Existing approaches can detect at most 9 of the 54 unique logic bugs detected by ERIQ, indicating that many ERIQ-detected bugs are distinctive and that ERIQ complements existing approaches. Under the 24-hour budget, ERIQ detects 54 bugs versus 16 for the best-performing baseline, demonstrating its stronger bug detection capability.}

\subsection{RQ3: Contribution of the Three Representations of Intermediate Query Results}\label{sec4.4} 
To answer RQ3, we analyzed how the VIEW, CTE, and TEMPT representations contribute to ERIQ's bug detection. For each of the 54 unique logic bugs, we classified its minimized test case according to the result inconsistency patterns defined in Section~\ref{sec3.4}. We then compared ERIQ's strategy of considering all three representations and comparing all available SQL variants with three pairwise strategies: VIEW--CTE, VIEW--TEMPT, and CTE--TEMPT. The inconsistency patterns reveal which variants return different results and therefore allow us to determine how many bugs each pairwise strategy could expose without identifying which result is incorrect.

\autoref{tab5} presents the distribution of the minimized test cases for the 54 unique logic bugs across the result inconsistency patterns. The pattern $\mathtt{V\ne C}$\texttt{(T\_SKIP)} indicates that the TEMPT variant cannot be legally constructed because the referencing query references the intermediate query result multiple times within the same statement, so ERIQ compares only the VIEW and CTE variants and observes inconsistent results. When all three variants return results, $\mathtt{V=T\ne C}$, $\mathtt{V=C\ne T}$, and $\mathtt{C=T\ne V}$ indicate that the CTE, TEMPT, and VIEW variants, respectively, return results different from the equal results of the other two variants. The pattern \texttt{all\_diff} indicates that all three results are pairwise different.

\begin{table}[htbp]
\centering
\caption{Distribution of the minimized test cases for the 54 unique logic bugs across result inconsistency patterns. V = VIEW, C = CTE, and T = TEMPT.}
\label{tab5}
\begin{tabular}{l@{\hskip 6pt}r@{\hskip 6pt}r@{\hskip 6pt}r@{\hskip 6pt}r@{\hskip 6pt}r@{\hskip 6pt}r}
\toprule
\textbf{DBMS} &
\textbf{\makecell{$\mathtt{V \ne C}$\\ \texttt{(T\_SKIP)}}} &
\textbf{\makecell{$\mathtt{V = T}$\\$\mathtt{\ne C}$}} &
\textbf{\makecell{$\mathtt{V = C}$\\$\mathtt{\ne T}$}} &
\textbf{\makecell{$\mathtt{C = T}$\\$\mathtt{\ne V}$}} &
\textbf{all\_diff} &
\textbf{Sum} \\
\midrule
MySQL       & 17 & 0 & 8 & 1 & 0 & 26 \\
MariaDB     & 0  & 0 & 9 & 1 & 1 & 11 \\
Percona     & 9  & 0 & 4 & 0 & 0 & 13 \\
OceanBase   & 2  & 0 & 2 & 0 & 0 & 4 \\
\midrule
\textbf{Total} & \textbf{28} & \textbf{0} & \textbf{23} & \textbf{2} & \textbf{1} & \textbf{54} \\
\bottomrule
\end{tabular}
\end{table}

The most frequent pattern is $\mathtt{V \ne C}$\texttt{(T\_SKIP)}, which accounts for 28 of the 54 test cases (51.9\%). These cases demonstrate the contribution of the VIEW--CTE comparison when the TEMPT variant is unavailable. The second most frequent pattern is $\mathtt{V=C\ne T}$, which accounts for 23 cases (42.6\%). Because the VIEW and CTE variants return identical results in these cases, the inconsistency can be exposed only when the TEMPT variant is included in the comparison. The remaining test cases include 2 instances (3.7\%) of $\mathtt{C=T\ne V}$ and 1 instance (1.9\%) of \texttt{all\_diff}. Although none of the 54 test cases exhibits the $\mathtt{V=T\ne C}$ pattern, ERIQ also generated a report with this pattern (\href{https://jira.mariadb.org/browse/MDEV-40246}{MDEV\#40246}). Developers classified this report as a duplicate of \href{https://jira.mariadb.org/browse/MDEV-40143}{MDEV\#40143}. The minimized test case for MDEV\#40143 used in this analysis exhibits the \texttt{all\_diff} pattern.

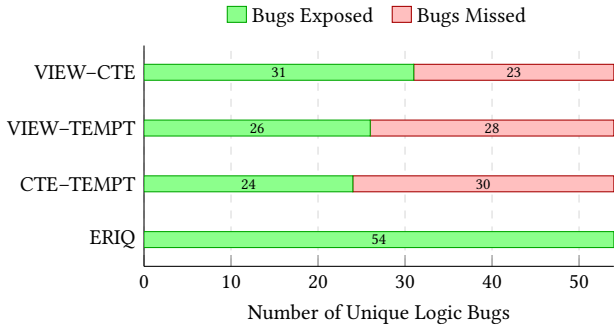
\begin{figure}[htbp]
\centering
\begin{tikzpicture}
\begin{axis}[
    xbar stacked,
    width=0.92\linewidth,
    height=4.5cm,
    bar width=6pt,
    xmin=0,
    xmax=54,
    xlabel={Number of Unique Logic Bugs},
    ytick={1,2,3,4},
    yticklabels={
        {ERIQ},
        CTE--TEMPT,
        VIEW--TEMPT,
        VIEW--CTE
    },
    enlarge y limits=0.16,
    axis x line*=bottom,
    axis y line*=left,
    xtick={0,10,20,30,40,50},
    xmajorgrids,
    grid style={dashed,gray!30},
    tick label style={font=\small},
    label style={font=\small},
    legend style={
        at={(0.5,1.04)},
        anchor=south,
        legend columns=2,
        draw=none,
        font=\small,
        /tikz/every even column/.append style={column sep=5pt}
    }
]

% Bugs Exposed
\addplot+[
    xbar,
    fill=green!45,
    draw=green!65!black,
    point meta=explicit symbolic,
    nodes near coords={\pgfplotspointmeta},
    every node near coord/.append style={
        font=\scriptsize,
        text=black
    }
]
coordinates {
    (54,1) [54]
    (24,2) [24]
    (26,3) [26]
    (31,4) [31]
};

% Bugs Missed
\addplot+[
    xbar,
    fill=red!25,
    draw=red!65!black,
    point meta=explicit symbolic,
    nodes near coords={\pgfplotspointmeta},
    every node near coord/.append style={
        font=\scriptsize,
        text=black
    }
]
coordinates {
    (0,1)  []
    (30,2) [30]
    (28,3) [28]
    (23,4) [23]
};

\legend{Bugs Exposed,Bugs Missed}

\end{axis}
\end{tikzpicture}

\caption{Comparison of the three pairwise comparison strategies and ERIQ based on the minimized test cases for the 54 unique logic bugs.}
\label{fig:pairwise}
\end{figure}

We next quantified the advantage of considering all three representations over each pairwise strategy. \autoref{fig:pairwise} summarizes the results.

Based on these test cases, restricting the comparison to VIEW--CTE, VIEW--TEMPT, or CTE--TEMPT would expose 31, 26, and 24 bugs out of 54, respectively, leaving 23, 28, and 30 bugs undetected. In contrast, ERIQ considers all three representations and compares the results of all available SQL variants for each test case, thereby exposing all 54 bugs.

\finding{The VIEW--CTE, VIEW--TEMPT, and CTE--TEMPT strategies expose 31, 26, and 24 bugs out of 54, respectively, while ERIQ exposes all 54 by considering the three representations, demonstrating their complementary contributions to ERIQ's bug detection.}

\subsection{RQ4: Bug Analysis}
\label{sec4.5}

To answer RQ4, we analyzed the SQL features involved in the minimized test cases for the 54 unique logic bugs and examined the root causes of representative bugs.
We first classified the minimized test cases according to the SQL features involved in triggering their result inconsistencies. We then selected five representative bugs for in-depth root-cause analysis.

\textbf{Analysis of SQL Features Triggering Result Inconsistencies.}
We manually inspected the minimized test case for each bug and classified the SQL features involved in triggering its result inconsistency. Because one test case may contain multiple SQL features, we used a multi-label classification and counted each bug in every applicable category. We considered seven SQL-feature categories: Group/Aggregate, Union, Subquery, Join, Distinct, Order/Limit, and Window. Group/Aggregate includes \texttt{GROUP BY} clauses, \texttt{HAVING} clauses, and aggregate function calls. Union includes \texttt{UNION} and \texttt{UNION ALL} operators. Subquery includes scalar subqueries, derived tables, and subqueries used with \texttt{EXISTS}, \texttt{NOT EXISTS}, \texttt{IN}, and \texttt{NOT IN}. Join includes explicit join operations such as \texttt{JOIN}, \texttt{LEFT JOIN}, and \texttt{RIGHT JOIN}. Distinct includes both \texttt{SELECT DISTINCT} clauses and aggregate functions qualified with \texttt{DISTINCT}. Order/Limit includes \texttt{ORDER BY} and \texttt{LIMIT} clauses. Window includes window-function calls containing an \texttt{OVER} clause. We counted each feature regardless of whether it appeared in the intermediate query or the referencing query. The categories are not mutually exclusive, and one SQL fragment may contribute to multiple categories. For example, \texttt{COUNT(DISTINCT ...)} is counted under both Group/Aggregate and Distinct, whereas \texttt{COUNT(*) OVER (...)} is counted under both Group/Aggregate and Window.

%\begin{table}[t]
\begin{table}[htbp]
\centering
\caption{SQL features involved in triggering result inconsistencies for the 54 unique logic bugs. A bug may involve multiple features.}
\label{tab:rq4_features}
\resizebox{\columnwidth}{!}{
\begin{tabular}{lrrrrr}
\toprule
\textbf{SQL Feature} &
\textbf{MySQL} &
\textbf{MariaDB} &
\textbf{Percona} &
\textbf{OceanBase} &
\textbf{Total} \\
\midrule
Group/Aggregate & 13 & 4 & 6 & 1 & 24 \\
Union           & 14 & 1 & 5 & 3 & 23 \\
Subquery        & 10 & 3 & 5 & 1 & 19 \\
Join            &  6 & 3 & 5 & 0 & 14 \\
Distinct        &  3 & 3 & 2 & 1 &  9 \\
Order/Limit     &  1 & 2 & 1 & 0 &  4 \\
Window          &  1 & 1 & 0 & 0 &  2 \\
\bottomrule
\end{tabular}}
\end{table}

\autoref{tab:rq4_features} shows that the test cases triggering result inconsistencies involve diverse SQL features. Group/Aggregate is the most frequent category, appearing in 24 of the 54 bugs (44.4\%), followed by Union in 23 bugs (42.6\%), Subquery in 19 bugs (35.2\%), and Join in 14 bugs (25.9\%). Distinct, Order/Limit, and Window appear in 9 bugs (16.7\%), 4 bugs (7.4\%), and 2 bugs (3.7\%), respectively. Overall, 47 of the 54 bugs (87.0\%) involve at least one of the seven categories, and 37 bugs (68.5\%) involve at least two categories. These results show that the bugs exposed by ERIQ involve a broad range of SQL features and that many involve multiple feature categories.

The minimized test cases for the remaining seven bugs do not involve any of these seven categories. Among them, four involve temporal functions, two involve JSON-related expressions, and one involves an \texttt{ENUM}-valued \texttt{CASE} expression compared with a numeric value. These cases show that ERIQ can also expose bugs involving functions and expressions outside the seven SQL-feature categories considered above.

\textbf{Root-Cause Analysis of Representative Bugs.}
To further understand why the detected bugs produce incorrect results, we selected five representative bugs spanning diverse root causes and for which developer feedback or fixes provided sufficient evidence for in-depth analysis. Their identified causes include defects in function implementation, expression evaluation, derived merging, \texttt{DISTINCT} handling, and type information loss during materialization.

\emph{Case 1: Incorrect state management across rows in the \texttt{ST\_IsValid} function.}
Listing~\ref{lst1} presents a MariaDB bug
(\href{https://jira.mariadb.org/browse/MDEV-40143}{MDEV\#40143})
in which the VIEW, CTE, and TEMPT variants return one row with value~\texttt{1}, one row with value~\texttt{NULL}, and two rows with value~\texttt{1}, respectively. Developer analysis confirmed that the root cause was a defect in \texttt{ST\_IsValid}, which retained stale nullability state across row evaluations. When \texttt{derived\_merge} is disabled, both the VIEW and CTE variants return one row with value~\texttt{1}. Thus, derived merging affects how the defect manifests but does not eliminate it, because both variants still differ from the expected two-row result. Together with the correct TEMPT result, this observation shows that the same function-level defect can manifest differently across optimization and materialization paths. After the fix, all three variants consistently returned two rows with value~\texttt{1}.

\emph{Case 2: Incorrect numeric evaluation of \texttt{DAYNAME} in a Boolean context.}
Listing~\ref{lst:bug_dayname} presents a MySQL bug
(\href{https://bugs.mysql.com/bug.php?id=120844}{MySQL\#120844})
in which the VIEW variant returns an empty result, whereas the CTE variant returns the two expected rows. Because \texttt{DAYNAME} returns a nonnumeric string, its conversion in a Boolean context should yield zero, leaving the inner subquery empty. The corresponding MySQL worklog shows that \texttt{Item\_func\_dayname} inherited from the integer-valued \texttt{Item\_func\_weekday}, causing \texttt{DAYNAME} to use an integer-valued evaluation interface and return a weekday index when evaluated numerically. MySQL fixed the defect by changing its base class to \texttt{Item\_str\_func}.

\begin{lstlisting}[
  float=h,
  caption={MySQL\#120844. Incorrect numeric evaluation of MySQL's \texttt{DAYNAME} in a Boolean context.},
  label={lst:bug_dayname},
  language=SQL,
  frame=single,
  basicstyle=\footnotesize\ttfamily, escapeinside={(@}{@)},
aboveskip=-1em,
  belowskip=-1em
]
CREATE TABLE t0 (c1 VARCHAR(1), c3 DATE);
INSERT INTO t0 VALUES ('x', '7056-10-15');

-- VIEW variant
CREATE VIEW v0 AS
  SELECT '' AS vc_0, DAYNAME(c3) AS vc_1 FROM t0
  UNION SELECT c1, DAYNAME(c3) FROM t0;
SELECT vc_0 FROM v0
  WHERE vc_0 NOT IN (
  SELECT vc_0 FROM v0 WHERE vc_1);
(@\textcolor{red}{-- Returns: empty set $\times$}@)

-- CTE variant
WITH CTE AS (
  SELECT '' AS vc_0, DAYNAME(c3) AS vc_1 FROM t0
  UNION SELECT c1, DAYNAME(c3) FROM t0)
SELECT vc_0 FROM CTE
  WHERE vc_0 NOT IN (
  SELECT vc_0 FROM CTE WHERE vc_1);
(@\textcolor{red}{-- Returns: '', 'x' $\checkmark$}@)
\end{lstlisting}

\emph{Case 3: Incorrect evaluation of an expression involving a \texttt{YEAR} column during derived result processing.}
Listing~\ref{lst:bug_derived_merge} presents a Percona bug
(\href{https://perconadev.atlassian.net/browse/PS-11410}{PS\#11410})
in which the VIEW and CTE variants return \texttt{999.9999}, whereas the TEMPT variant returns the expected value~\texttt{1063.5000}. Because \texttt{COALESCE(b.c1, 0)} evaluates to \texttt{2127}, dividing it by 2 should yield \texttt{1063.5000}. Developer analysis confirmed that the incorrect result was caused by incorrect expression evaluation during derived merging. The VIEW and CTE variants follow the derived-merging path, whereas the TEMPT variant materializes the intermediate query result before the subsequent computation and therefore does not exhibit this incorrect evaluation.

\begin{lstlisting}[
  float=h,
  caption={PS\#11410. Incorrect evaluation of an expression involving a \texttt{YEAR} column during derived result processing in Percona.},
  label={lst:bug_derived_merge},
  language=SQL,
  frame=single,
  basicstyle=\footnotesize\ttfamily, escapeinside={(@}{@)},
aboveskip=0.5em,
  belowskip=-1em
]
CREATE TABLE t0 (c1 YEAR);
INSERT INTO t0 VALUES (2127);

-- VIEW variant
CREATE VIEW v0 AS SELECT COALESCE(b.c1, 0) AS vc_1
  FROM t0 AS a JOIN t0 AS b;
SELECT vc_1 / 2 AS r FROM v0 ORDER BY 1;
(@\textcolor{red}{-- Returns: 999.9999 $\times$}@)

-- CTE variant
WITH CTE AS (SELECT COALESCE(b.c1, 0) AS vc_1
  FROM t0 AS a JOIN t0 AS b)
SELECT vc_1 / 2 AS r FROM CTE ORDER BY 1;
(@\textcolor{red}{-- Returns: 999.9999 $\times$}@)

-- TEMPT variant
CREATE TEMPORARY TABLE temp_t AS
  SELECT COALESCE(b.c1, 0) AS vc_1
FROM t0 AS a JOIN t0 AS b;
SELECT vc_1 / 2 AS r FROM temp_t ORDER BY 1;
(@\textcolor{red}{-- Returns: 1063.5000 $\checkmark$}@)
\end{lstlisting}

\emph{Case 4: Incorrect result for multi-column \texttt{DISTINCT} with \texttt{ORDER BY} over a \texttt{VIEW}.}
Listing~\ref{lst:bug_distinct_view} presents a MariaDB bug
(\href{https://jira.mariadb.org/browse/MDEV-40245}{MDEV\#40245})
in which the VIEW variant returns only one row, whereas the CTE and TEMPT variants return the two expected rows. Because \texttt{NULLIF(1, 0)} evaluates to \texttt{1}, the expression \texttt{c0 / NULLIF(1, 0)} evaluates to \texttt{2.0000} and \texttt{4.0000}, respectively. The resulting rows are therefore distinct and should both be preserved by \texttt{DISTINCT}. Developer analysis identified the cause as incorrect \texttt{DISTINCT} handling in a multi-column \texttt{DISTINCT} query with \texttt{ORDER BY} over a \texttt{VIEW}, where one projected column is constant-valued and another projected expression contains a constant-foldable subexpression. In this case, the VIEW variant fails to preserve both distinct rows, whereas the equivalent CTE and TEMPT variants return the expected results.

\begin{lstlisting}[
  float=htbp,
  caption={MDEV\#40245. Incorrect result for multi-column \texttt{DISTINCT} with \texttt{ORDER BY} over a \texttt{VIEW} in MariaDB.},
  label={lst:bug_distinct_view},
  language=SQL,
  frame=single,
  basicstyle=\footnotesize\ttfamily,
  escapeinside={(@}{@)},
aboveskip=-1em,
  belowskip=-1em
]
CREATE TABLE t0 (c0 INT, c1 INT);
INSERT INTO t0 VALUES (2, NULL), (4, NULL);

-- VIEW variant
CREATE VIEW v0 AS SELECT DISTINCT c1 AS vc_0,
  c0 / NULLIF(1, 0) AS vc_1 FROM t0 ORDER BY vc_0, vc_1;
SELECT * FROM v0;
(@\textcolor{red}{-- Returns: (NULL, 2.0000)$\times$}@)

-- CTE variant
WITH CTE AS (SELECT DISTINCT c1 AS vc_0,
  c0 / NULLIF(1, 0) AS vc_1 FROM t0 ORDER BY vc_0, vc_1)
SELECT * FROM CTE;
(@\textcolor{red}{-- Returns: (NULL, 2.0000), (NULL, 4.0000)$\checkmark$}@)

-- TEMPT variant
CREATE TEMPORARY TABLE temp_t AS 
  SELECT DISTINCT c1 AS vc_0,
  c0 / NULLIF(1, 0) AS vc_1 FROM t0 ORDER BY vc_0, vc_1;
SELECT * FROM temp_t;
(@\textcolor{red}{-- Returns: (NULL, 2.0000), (NULL, 4.0000)$\checkmark$}@)
\end{lstlisting}

\emph{Case 5: Loss of JSON Boolean type information during materialization.}
Listing~\ref{lst:bug_json_boolean} presents a MariaDB bug
(\href{https://jira.mariadb.org/browse/MDEV-40252}{MDEV\#40252})
in which the VIEW and CTE variants return one row with value~\texttt{true}, whereas the TEMPT variant returns an empty result. Developer analysis confirmed that the defect occurs during the materialization of the JSON Boolean value returned by \texttt{JSON\_EXTRACT}. During materialization, the JSON Boolean loses its original type information and is subsequently treated as a regular string. Therefore, when the predicate \texttt{vc\_1 OR 0} is evaluated, MariaDB applies string-to-number conversion, causing \texttt{true} to be converted to \texttt{0}. As a result, the predicate evaluates to false and incorrectly filters out the row in the TEMPT variant.

\begin{lstlisting}[
  float=h,
  caption={MDEV\#40252. Loss of JSON Boolean type information during materialization in MariaDB.},
  label={lst:bug_json_boolean},
  language=SQL,
  frame=single,
  basicstyle=\footnotesize\ttfamily,
  escapeinside={(@}{@)},
aboveskip=0.5em,
  belowskip=-1em
]
CREATE TABLE t0 (c2 JSON);
INSERT INTO t0 VALUES ('true');

-- VIEW variant
CREATE VIEW v0 AS
  SELECT JSON_EXTRACT(c2, '$') AS vc_1 FROM t0;
SELECT * FROM v0 WHERE vc_1 OR 0;
(@\textcolor{red}{-- Returns: true $\checkmark$}@)

-- CTE variant
WITH CTE AS (
  SELECT JSON_EXTRACT(c2, '$') AS vc_1 FROM t0)
SELECT * FROM CTE WHERE vc_1 OR 0;
(@\textcolor{red}{-- Returns: true $\checkmark$}@)

-- TEMPT variant
CREATE TEMPORARY TABLE temp_t AS
  SELECT JSON_EXTRACT(c2, '$') AS vc_1 FROM t0;
SELECT * FROM temp_t WHERE vc_1 OR 0;
(@\textcolor{red}{-- Returns: empty set $\times$}@)
\end{lstlisting}

\finding{The minimized test cases for the 54 unique logic bugs involve diverse SQL features, and the five representative bugs exhibit varied root causes, showing that ERIQ can expose DBMS logic bugs with diverse characteristics.}

\section{Discussion}\label{sec5}
\textbf{False Positive Mitigation.}
ERIQ takes several measures to mitigate false positives caused by the testing procedure itself. It excludes nondeterministic functions during query generation, thereby avoiding false positives arising from nondeterministic evaluations. ERIQ also compares only legally constructed SQL variants; when an equivalent TEMPT variant cannot be constructed under the current strategy, ERIQ skips the TEMPT variant and compares only the VIEW and CTE variants. In addition, ERIQ normalizes returned values and compares them using multiset semantics to mitigate inconsistencies caused solely by result representation or unspecified row ordering. In our evaluation, developers confirmed 63 of the 64 bugs; among them, 62 corresponded to logic bugs (54 unique and 8 duplicates), while one was classified as the documentation issue discussed below. None of the 63 confirmed bugs was identified as a false positive.

\textbf{Documentation Issue.}
One of the 64 bugs, \href{https://bugs.mysql.com/bug.php?id=120856}{MySQL\#120856}, was confirmed by MySQL developers and classified as a documentation issue. As shown in Listing~\ref{lst:doc_issue}, the VIEW and CTE variants return \texttt{4} for \texttt{COERCIBILITY(vc)}, whereas the TEMPT variant returns \texttt{2}. The MySQL developer explained that, in the VIEW and CTE variants, \texttt{BIN(c2)} is a character-string function and therefore has coercibility \texttt{COERCIBLE} (\texttt{4}). In contrast, after its result is materialized by \texttt{CREATE TEMPORARY TABLE ... AS SELECT}, it becomes a table column with coercibility \texttt{IMPLICIT} (\texttt{2})~\cite{mysql_coercibility}. The developer classified the report as a documentation issue because the coercibility of string functions was not clearly specified in the official documentation. Accordingly, we do not count this report as a logic bug; instead, it reveals a gap in MySQL's documentation concerning the coercibility of string functions.

\begin{lstlisting}[
  float=h,
  caption={MySQL\#120856. Different \texttt{COERCIBILITY} results between expression-derived and materialized values in MySQL.},
  label={lst:doc_issue},
  language=SQL,
  frame=single,
  basicstyle=\footnotesize\ttfamily,
  escapeinside={(@}{@)},
aboveskip=0.5em,
  belowskip=-1em
]
CREATE TABLE t0 (c2 BIGINT);
INSERT INTO t0 VALUES (100);

-- VIEW variant
CREATE VIEW v0 AS
  SELECT BIN(c2) AS vc FROM t0;
SELECT COERCIBILITY(vc) FROM v0;
(@\textcolor{red}{-- Returns: 4}@)

-- CTE variant
WITH CTE AS (
  SELECT BIN(c2) AS vc FROM t0)
SELECT COERCIBILITY(vc) FROM CTE;
(@\textcolor{red}{-- Returns: 4}@)

-- TEMPT variant
CREATE TEMPORARY TABLE temp_t AS
  SELECT BIN(c2) AS vc FROM t0;
SELECT COERCIBILITY(vc) FROM temp_t;
(@\textcolor{red}{-- Returns: 2}@)
\end{lstlisting}

\textbf{Limitations.}
ERIQ has several limitations resulting from its equivalence requirements and result-comparison strategy. First, nondeterministic functions are excluded to avoid inconsistent results caused by different evaluations, so bugs in such functions cannot currently be detected. Second, ERIQ does not generate recursive CTEs because the recursive self-references required by such CTEs cannot be represented by the equivalent VIEW and TEMPT forms currently constructed by ERIQ. Third, when an equivalent TEMPT variant cannot be legally constructed, ERIQ can compare only the VIEW and CTE variants, potentially missing bugs that require TEMPT for exposure. Finally, ERIQ compares results using multiset semantics and therefore does not target bugs that manifest solely as differences in row ordering.

\section{Related Work}\label{sec6}
\textbf{DBMS Logic Bug Detection.} Existing approaches have proposed various techniques for detecting DBMS logic bugs~\cite{sqlancer,pqs,norec,tlp,pinolo,eet,coddtest,tqs1,tqs2,dqp,radar,ddlcheck,edc}. PQS~\cite{pqs} synthesizes queries that are guaranteed to return a selected pivot row and detects bugs when the pivot row is missing from the result. NoREC~\cite{norec} rewrites a query into a form that inhibits DBMS optimizations and compares the results of the original and rewritten queries. TLP~\cite{tlp} partitions a query according to ternary logic and checks whether the combined results of the partitioned queries are consistent with the original query result. Pinolo~\cite{pinolo} synthesizes queries whose results should be supersets or subsets of a seed query result and detects violations of the expected containment relation. EET~\cite{eet} and CODDTest~\cite{coddtest} apply equivalent transformations to query expressions and compare the results of the original and transformed queries. Radar~\cite{radar} compares the results of the same query on databases containing the same data but different metadata. EDC~\cite{edc} replaces data operation expressions with precomputed results and compares the original and transformed query results.

In contrast to these approaches, ERIQ checks result consistency across equivalent representations of intermediate query results. It constructs SQL variants that use a VIEW, a CTE, or a TEMPT to represent the same intermediate query result and checks whether the variants return consistent results. ERIQ therefore explores a new perspective for detecting DBMS logic bugs, complementing existing approaches.

\textbf{DBMS Test Case Generation.} Existing DBMS test case generation techniques aim to improve the validity and diversity of generated SQL inputs~\cite{adusa,ratel,sedar,slutz,sqlsmith,squirrel,griffin,lego,sqlright,dynsql,qpg,mist,dbms_survey}. SQLsmith~\cite{sqlsmith} randomly constructs queries using database-schema information and AST-based rules. SQUIRREL~\cite{squirrel} performs type-based mutations on an intermediate representation and uses dependency information to improve the syntactic and semantic validity of generated statements. Griffin~\cite{griffin} constructs metadata graphs that capture dependencies between statements and database objects and uses them for statement reshuffling and metadata-guided substitution. LEGO~\cite{lego} uses coverage feedback to discover SQL type affinities and progressively generate diverse SQL statement sequences. SQLRight~\cite{sqlright} combines coverage guidance with validity-oriented mutations, whereas DynSQL~\cite{dynsql} dynamically collects database-state information to incrementally construct complex and valid SQL inputs. QPG~\cite{qpg} uses query-plan diversity to guide database-state mutations, and MIST~\cite{mist} combines hierarchical SQL-feature guidance, error feedback, and coverage-guided Monte Carlo tree search for LLM-based DBMS test-case generation. 

ERIQ complements these approaches by constructing equivalent representations of intermediate query results and checking their result consistency. The diverse query-generation techniques employed by existing approaches could be incorporated into ERIQ to generate more diverse intermediate queries and referencing queries, potentially exposing additional DBMS logic bugs. Conversely, ERIQ's representation-based consistency checking could be integrated with these generation techniques to further enhance their bug detection capability.

\section{Conclusion}\label{sec7}
This paper presents ERIQ, a novel testing approach for detecting DBMS logic bugs by checking result consistency across equivalent representations of intermediate query results. ERIQ constructs SQL variants that use a VIEW, a CTE, or a TEMPT to represent the same intermediate query result and compares their returned results. By comparing results across these equivalent representations, ERIQ introduces a new perspective on DBMS logic bug detection that complements existing approaches. We evaluated ERIQ on MySQL, MariaDB, Percona, and OceanBase. In total, ERIQ detected 64 bugs, 63 of which were confirmed by developers, and two have been fixed. Among the confirmed bugs, 54 were unique and previously unknown logic bugs, and one was a documentation issue. In future work, we plan to extend ERIQ to additional DBMSs, such as PostgreSQL, TiDB, and SQLite, and explore additional mechanisms for representing intermediate query results, such as derived tables, materialized views where supported, and persistent tables created from query results.

\begin{acks}
This work was supported by the Key Program of the National Natural Science Foundation of China under Grant No. 62532008.
\end{acks}

\bibliographystyle{ACM-Reference-Format}
\bibliography{sample}

\end{document}